\documentclass[a4paper,10pt,twoside]{cpc-hepnp}
\usepackage{CJK,upgreek,fancyhdr}
\usepackage{multicol}
\usepackage{graphicx}
\usepackage{booktabs}
\usepackage{amssymb,bm,mathrsfs,bbm,amscd}
\usepackage[tbtags]{amsmath}
\usepackage{lastpage}
\usepackage{subfigure}

\begin{document}
\begin{CJK*}{GB}{gbsn}

\fancyhead[c]{\small Chinese Physics C~~~Vol. xx, No. x (201x) xxxxxx}
\fancyfoot[C]{\small 010201-\thepage}

\footnotetext[0]{Received 31 June 2015}

\title{Self-Seeded FEL Wavelength Extension with High-Gain Harmonic Generation}

\author{
      Ling Zeng(ÔøÁè) $^{1}$
\quad Weilun Qin(ÇØΰÂ×)$^{1}$
\quad Gang Zhao (ÕÔ¸Õ)$^{1}$
\quad Senlin Huang (»ÆÉ­ÁÖ)$^{1;1)}$\email{huangsl@pku.edu.cn}%
\quad  \\Yuantao Ding $^{2}$
\quad Zhirong Huang $^{2}$
\quad Gabriel Marcus $^{2}$
\quad Kexin Liu(Áõ¿ËÐÂ)$^{1}$
}

\maketitle

\address{%

$^1$ Institute of Heavy Ion Physics, School of Physics, Peking University, Beijing 100871, China \\
$^2$ SLAC National Accelerator Laboratory, Menlo Park, CA 94025, USA\\

}

\begin{abstract}
We study a self-seeded high-gain harmonic generation (HGHG) free-electron laser (FEL) scheme to extend the wavelength of a soft X-ray FEL. This scheme uses a regular self-seeding monochromator to generate a seed laser at the wavelength of 1.52 nm, followed by a HGHG configuration to produce coherent, narrow-bandwidth harmonic radiations at the GW level. The 2nd and 3rd harmonic radiation are investigated with start-to-end simulations. Detailed studies on the FEL performance and shot-to-shot fluctuations are presented.
\end{abstract}



\footnotetext[0]{\hspace*{-3mm}\raisebox{0.3ex}{$\scriptstyle\copyright$}2013
Chinese Physical Society and the Institute of High Energy Physics
of the Chinese Academy of Sciences and the Institute
of Modern Physics of the Chinese Academy of Sciences and IOP Publishing Ltd}%

\begin{multicols}{2}

\section{Introduction}

X-ray free electron lasers (FELs), which demonstrate an improvement in peak brightness of approximately ten orders of magnitude over third-generation light source, have shown remarkable scientific capabilities in chemistry, biology, material science, as well as many other disciplines. There are two main schemes for single pass short wavelength FELs: SASE~\cite{SLAC-R-593,DESY097} and HGHG~\cite{3,4}. Until recently, most of the modern high-gain FELs in short wavelength (e.g., X-ray) region have been operated in SASE mode, such as the LCLS and the SACLA FEL \cite{5,6}, which is characterized by excellent transverse coherence. However, SASE FEL has poor temporal coherence and large shot-to-shot fluctuations in both the time and frequency domain since it starts from shot noise~\cite{zrh_kjkim_2007}.

HGHG FEL can generate fully coherent, high gain harmonic radiation of a seed laser. However, the harmonic number ($n$) of a single-stage HGHG FEL is limited by the requirement that the induced energy spread be less than the Pierce parameter ($\rho$) in the radiation undulator (radiator) to achieve high gain. So far, the highest harmonic obtained with single-stage HGHG is the 13th harmonic at 20 nm using a 1.2 GeV electron beam at FERMI~\cite{7}. In order to reach higher harmonics, so as to obtain shorter wavelength fully coherent FEL, several schemes have been proposed in recent years. Among them, the cascaded HGHG scheme with the help of ``fresh bunch'' technique was proposed in 2001~\cite{8}. Recently, 4.3 nm radiation (60th harmonic of a 260 nm UV seed laser) has been achieved with a two-stage HGHG configuration at FERMI~\cite{9}. One other scheme, EEHG~\cite{10}, was first proof-of-principle demonstrated at SLAC~\cite{11}. In 2011, researchers from Shanghai Institute of Applied Physics (SINAP) also observed the third harmonic from EEHG, which was further amplified to saturation~\cite{12}. Currently, EEHG at 160 nm (15th harmonic of a 2400 nm seed laser) has been produced at SLAC~\cite{13}. However, the cascaded HGHG and EEHG still have difficulty in generating hard X-ray FEL due to a lack of external seeds at X-ray wavelengths~\cite{14}.

To solve the difficulty of external seeding at very short wavelengths, DESY colleagues proposed an approach of self seeding in 1997~\cite{15}, and recently a simplified monochromator version for hard X-rays~\cite{HXSS}. This self-seeded FEL starts from a SASE stage, which operates in the linear regime. A following monochromator is used to generate a purified seed from the SASE radiation, and meanwhile the electron beam after the SASE stage goes through a bypass chicane. They recombine in an amplification undulator (amplifier stage) for further interaction, where the seed radiation gets amplified to saturation, producing near Fourier transform limited X-ray. The self-seeding approach works for both soft and hard X-ray FELs and has been successfully demonstrated recently~\cite{14,16}. It is worth noticing that two different configurations of monochromator have been used depending on the spectral range. For X-ray FEL with the photon energy below 2 keV, a grating-based monochromator has been used~\cite{14}, while for X-ray FEL with the photon energy above 4.5 keV, a diamond-based monochromator is more popular~\cite{16}. Within the photon energy range from 2 keV to 4.5 keV, the self-seeded FEL is difficult due to a lack of monochromator materials. This motivates to study alternative schemes to cover the energy gap 2-4.5 keV.

In this paper, we study a new scheme combining the self-seeding approach with HGHG to produce fully coherent X-rays.  It can not only fill the above photon energy gap not easily achieved by regular self-seeded FELs, but also extend the wavelength of a soft X-ray FEL machine to harder X-ray region. This self-seeded HGHG scheme will be described in Section~$\ref{sec:scheme}$, followed by the FEL simulation results in Section~$\ref{sec:simulation:long-mod}$ and ~$\ref{sec:simulation:fresh-bun}$.

\section{Self-seeded HGHG Scheme}\label{sec:scheme}

\begin{figure}[h]
\centering
\includegraphics[width=0.68\textwidth]{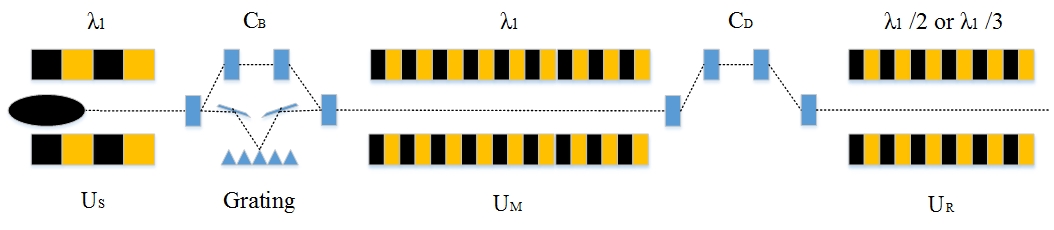}
\caption{\label{fig:config}   Schematic of self-seeded HGHG FEL. $U_S$ is a SASE undulator, $U_M$ is a long modulation undulator with combined functions as seed amplifier and HGHG modulator,  and $U_R$ is a radiation undulator (radiator) of HGHG. $C_B$ is a bypass chicane steering the electron beam around the X-ray monochromator, while $C_D$ is a dispersion chicane of HGHG.}
\end{figure}

The proposed setup of self-seeded HGHG scheme is shown in Fig. ~\ref{fig:config}, which consists of two stages: SASE stage and HGHG stage. The SASE stage follows the regular self-seeding configuration, comprising a SASE undulator, an X-ray monochromator, and an electron beam bypass chicane allowing room for the monochromator. An electron beam first traverses the SASE undulator, generating SASE radiation in the linear regime. After the SASE undulator, the radiation goes through the X-ray monochromator, which transmits a narrow-band of wavelength. The transmitted radiation will then be used as a seed for the following FEL amplifier. Meanwhile, the electron beam from the SASE undulator goes through a bypass chicane, being properly delayed, and recombines with the seed radiation at the entrance of HGHG stage. The bypass chicane also helps to wash out the microbunching of electron beam built up in the SASE undulator.

\end{multicols}
\begin{multicols}{2}

Compared to the external seed laser in the regular HGHG scheme, the seed radiation from the SASE stage has a much lower power, limited to a few hundred kilowatts herein considering the damaging threshold of the state-of-the-art X-ray monochromator optics~\cite{14}. As a result, we need to use a long modulation undulator with two combined functions. The first one is to amplify the seed radiation from the SASE stage, which is mainly achieved with the upstream part of the modulation undulator. The second one is to introduce energy modulation to the electron beam, as in a normal HGHG modulator. The energy modulated electron beam then goes through the dispersion chicane with proper $R_{56}$, getting density modulated, and radiates at the harmonic wavelength of the seed.

In this self-seeded HGHG scheme,  the electron beam quality degrades inevitably when it is used to amplify the seed radiation from the SASE stage. Therefore a compromise should be made between the modulation radiation power and the induced electron energy spread growth in the HGHG modulator.  In this paper, the modulation laser power is kept at hundred megawatt level to avoid a significant energy spread degradation of the electron beam in the seed amplification process.

\begin{figure}[h]
\centering
\includegraphics[width=0.75\textwidth]{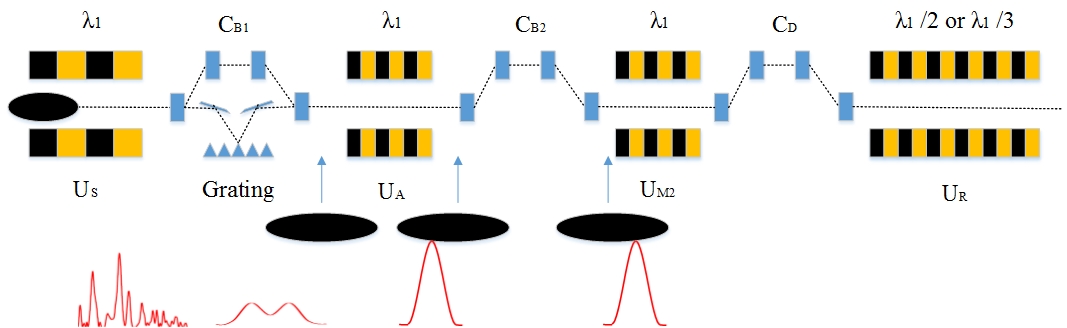}
\caption{\label{fig:config:2}   Self-seeded HGHG FEL with separated seed amplifier (using $U_A$ undulator) and modulator (using $U_{M2}$ undulator). $C_{B2}$ is an electron delay chicane.}
\end{figure}

To eliminate the impact of electron energy spread degradation in the seed amplification process, we have also proposed a self-seeded HGHG FEL setup with separated seed amplifier and modulator (see Fig.~\ref{fig:config:2}). In this case an electron beam with longer bunch length is used, which generates double-spike seed after the X-ray monochromator. The head spike of the seed is then aligned with the tail part of the electron bunch at the entrance of the amplifying undulator ($U_A$). Therefore only the tail part of the electron bunch is used to amplify the seed while the head part is kept undisturbed and ``fresh''. After the $U_A$ undulator, the electron bunch is delayed by a small chicane ($C_{B2}$), and consequently the head part is aligned with the seed radiation in the modulation undulator ($U_{M2}$) and gets energy-modulated. After the $U_{M2}$ undulator, the electron beam undergoes the same procedure as the above setup.

In the following discussions, the first setup is referred to as the long modulator case, while the second setup is referred to as the ``fresh bunch'' case.

\end{multicols}
\begin{multicols}{2}

\section{FEL simulation of long modulator case}\label{sec:simulation:long-mod}

As a representative example, we use the soft X-ray self-seeded (SXSS) FEL at LCLS to illustrate the feasibility of this scheme. Parameters are assumed based on the SXSS FEL for time-dependent FEL simulation using GENESIS~\cite{17} code. The electron beam has a central energy of 4.3 GeV, an uncorrelated energy spread of 1.0 MeV, and a normalized transverse emittance of 0.5 mm-mrad. It has a uniform current profile with a pulse duration of 30 fs and peak current of 2.5 kA.
The SASE undulator ($U_S$) is resonant at 1.52 nm. It uses 5 LCLS undulator segments and has a total length of 19.8 m (including the focusing optics in between). This is based on a consideration of keeping the SASE FEL power at a highest level while avoiding damage to the X-ray monochromator optics. The monochromator is assumed to have a Gaussian spectral response with a maximum power efficiency of 0.02 at 1.52 nm. An example\begin{center}
\includegraphics[width=0.22\textwidth]{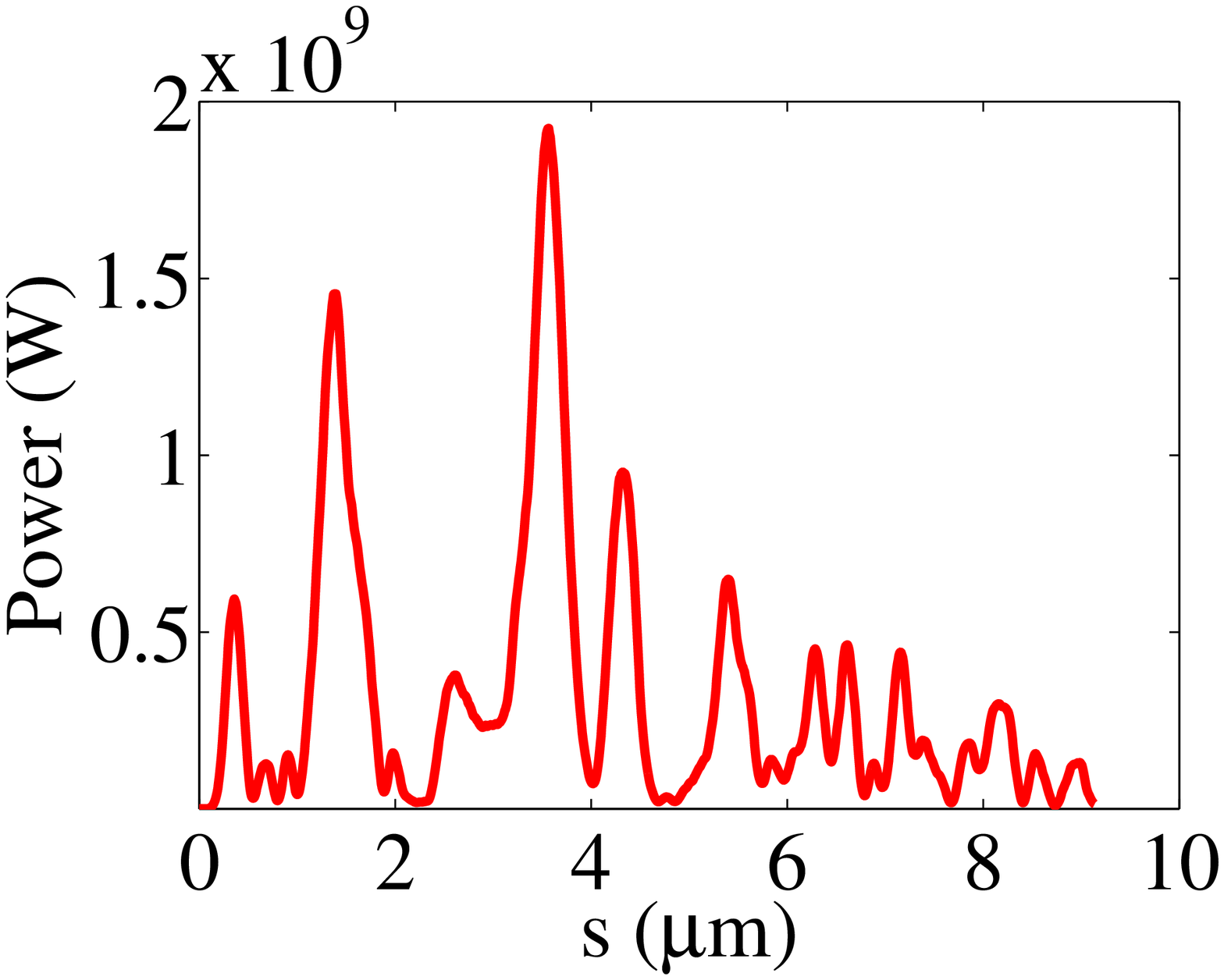}
\includegraphics[width=0.22\textwidth]{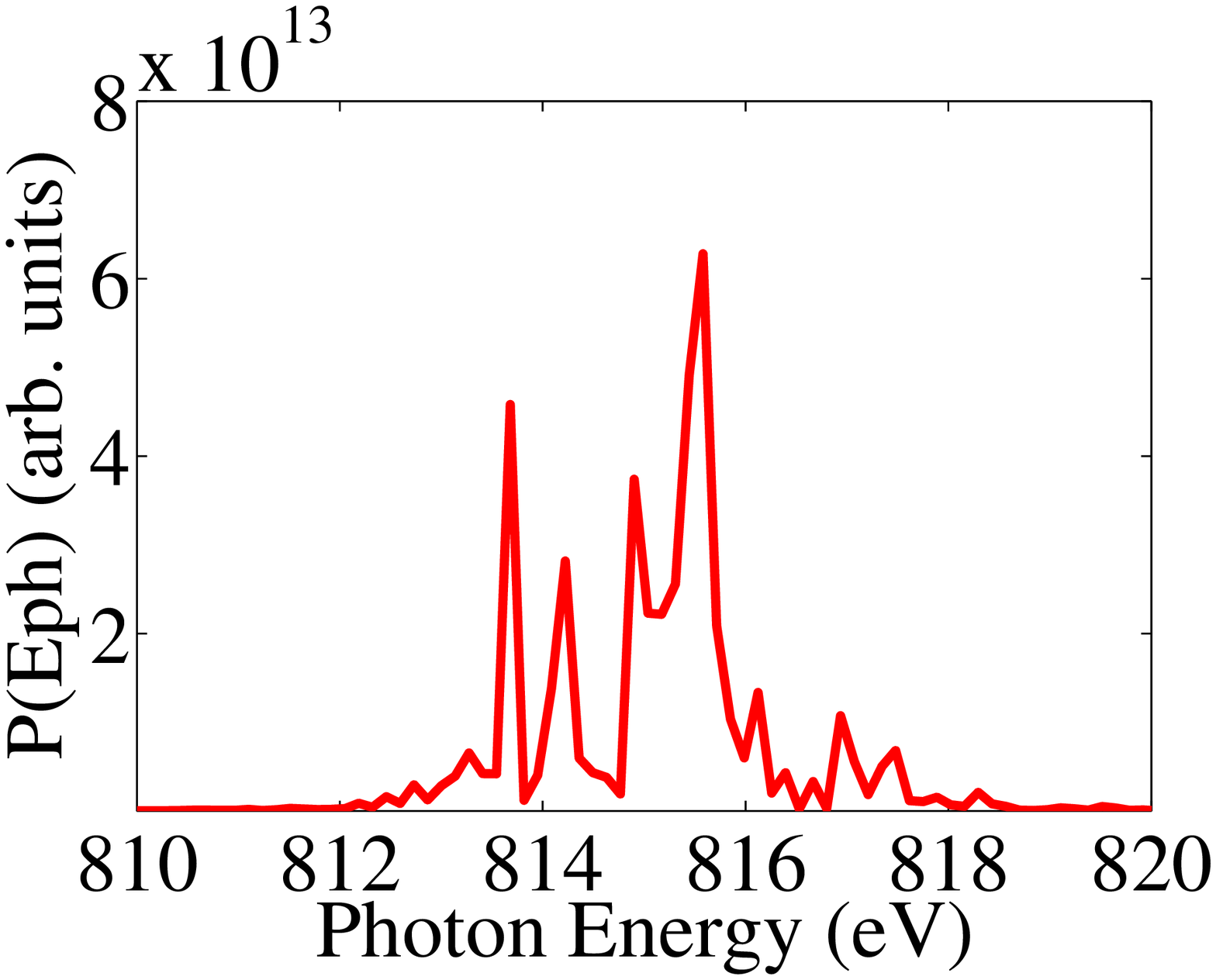}\\
\includegraphics[width=0.22\textwidth]{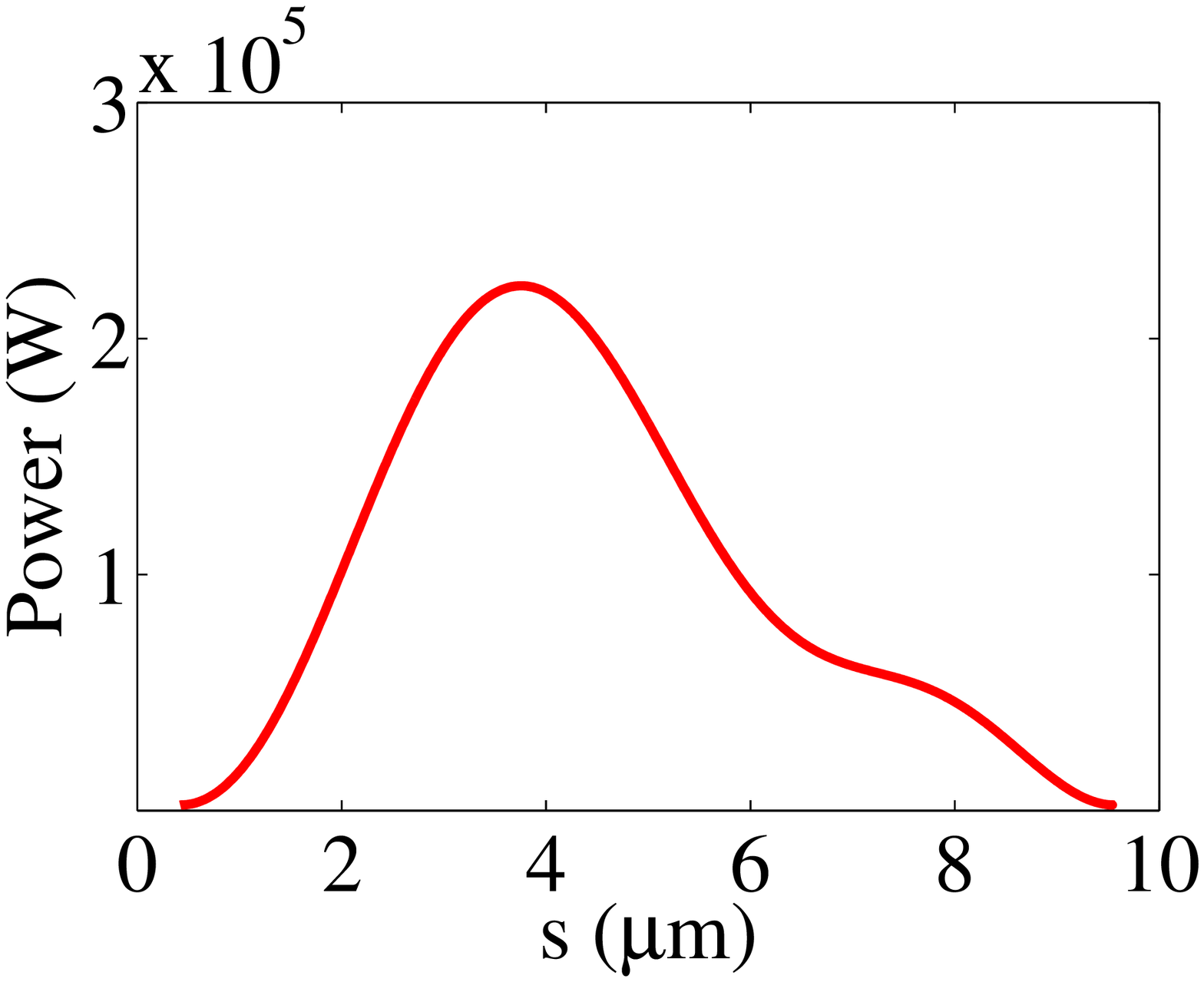}
\includegraphics[width=0.22\textwidth]{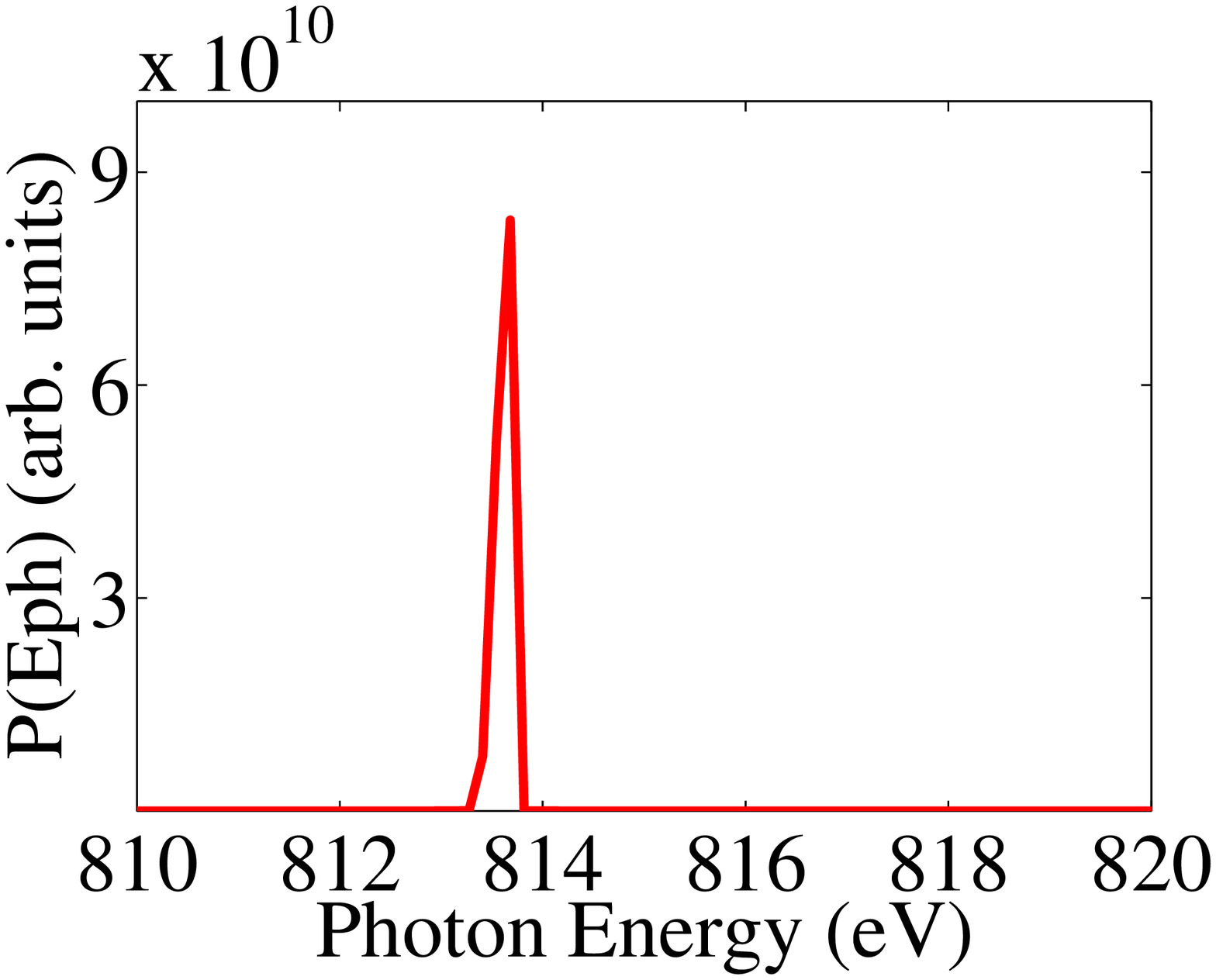}
\figcaption{\label{fig:seed:1} FEL power profiles (left) and spectra (right) at the exit of $U_S$ undulator (top) and after monochromator (bottom) in the long modulator case.}
\end{center} of the FEL power profiles and power spectra after the SASE undulator and X-ray monochromator is shown in Fig.~\ref{fig:seed:1}. One can see from the figure that after monochromator the radiation peak power drops to about 220 kW while a narrow bandwidth is filtered out.

\begin{center}
\includegraphics[width=0.40\textwidth]{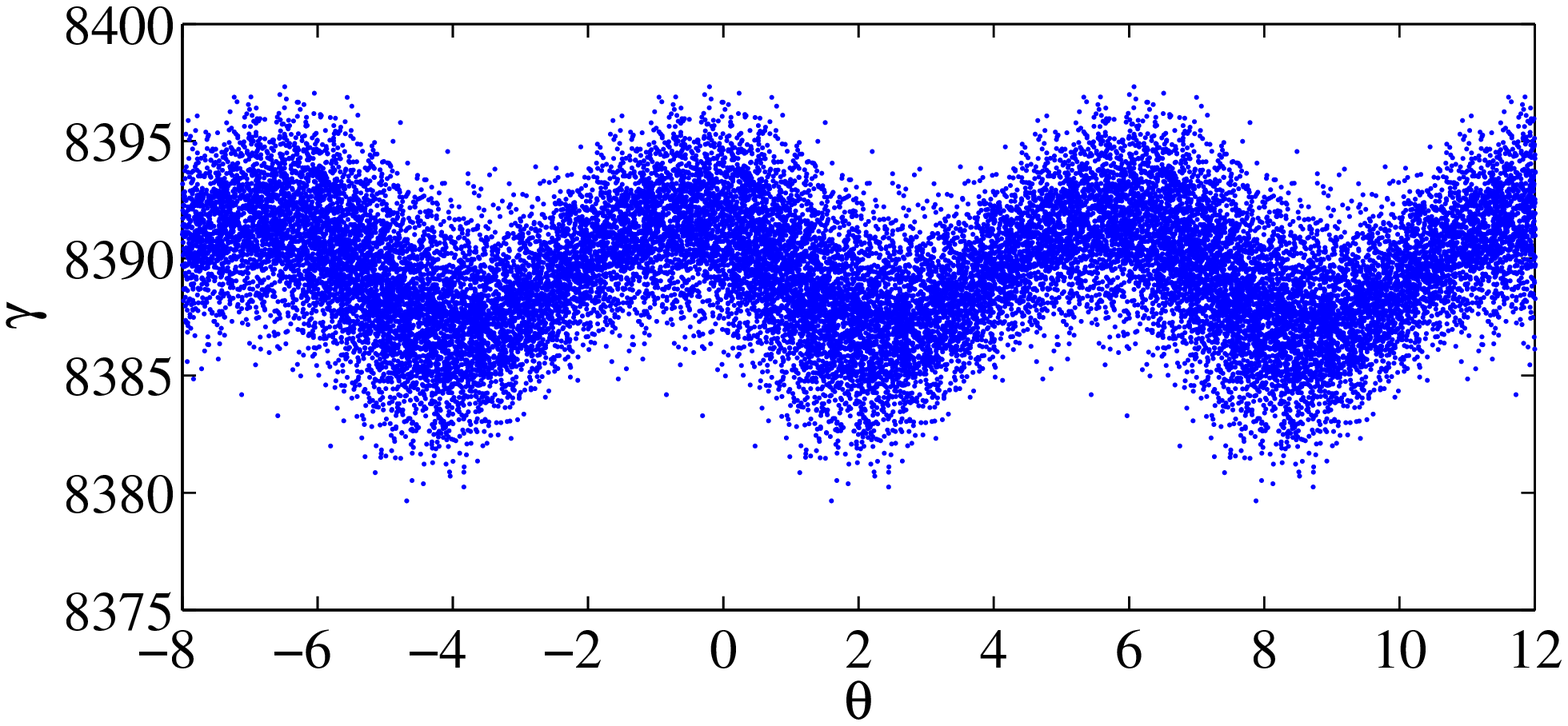}\\
\includegraphics[width=0.40\textwidth]{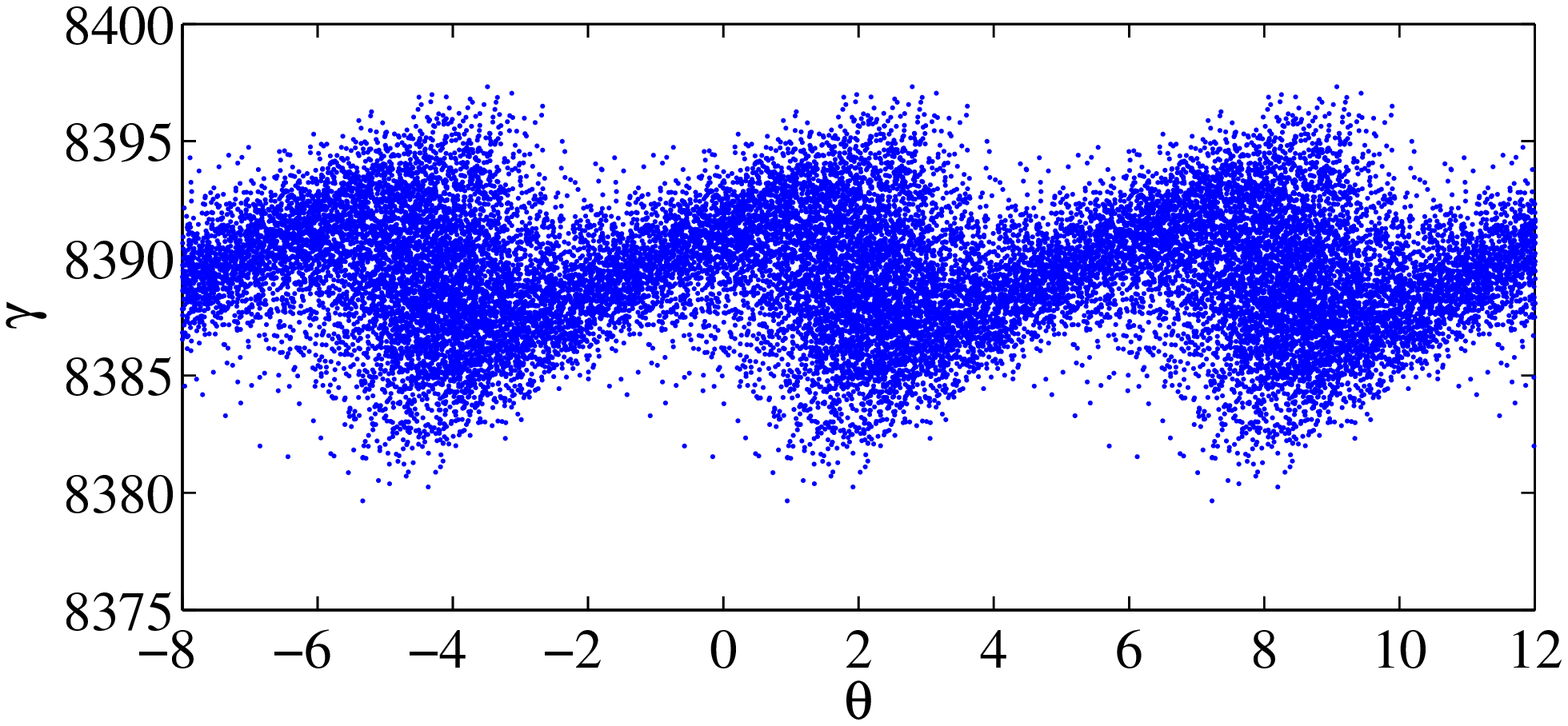}
\figcaption{\label{fig:long-phaspace}   Longitudinal phase space evolution in HGHG based on self-seeding scheme. (a) At the exit of modulator $U_M$; (b) At the exit of dispersion section $C_D$.}
\end{center}

For HGHG modulator, an optimal value for the electron energy modulation amplitude $\Delta\eta_m$ is ~\cite{3,4}
\begin{equation}
\Delta\eta_m\approx n\sigma_\eta
\label{eq:opt-mod-amp}
\end{equation}
with $\sigma_\eta$  the intrinsic uncorrelated energy spread and $n$ is the harmonic number. In order to obtain the optimal energy modulation, the length of $U_M$ undulator  needs to optimized. Herein we study the 2nd and 3rd harmonic generation and accordingly the undulator length is chosen to be 13 m and 14 m, respectively, to meet the above criteria of optimized energy modulation amplitude.

The momentum compaction factor $R_{56}$ of the chicane $C_D$ is set to satisfy~\cite{10}
\begin{equation}
R_{56}\Delta\gamma_m/\gamma\approx\lambda/4,
\label{eq:R56}
\end{equation}
where $\lambda$ is the wavelength of the seed laser. In this long modulator case, the value of $R_{56}$ are 0.56 $\mu m$ and 0.46 $\mu m$ for the 2nd and 3rd harmonic generation, respectively. The longitudinal phase space of electron beam after the modulation undulator ($U_M$) and dispersion chicane are shown in Fig.~\ref{fig:long-phaspace}.

The main electron and radiation parameters used in this study are summarized in Table $\ref{tab:simu-para}$. With these parameters, GENESIS simulations were performed. Figure~\ref{fig:fel:pow-vs-z} illustrates the evolution of the FEL power of the 2nd and 3rd harmonic radiation along the radiation undulator. We can see from the figure that the 2nd harmonic radiation at the exit of the radiation undulator is about 3.6 GW, while the 3rd harmonic radiation is about 0.9 GW. Figure~\ref{fig:fel:pas:1} shows the output power profile and spectra of the 2nd and 3rd harmonic radiation at the exit of the radiation undulator, which indicate that the normalized spectral width (FWHM) of the 2nd and 3rd harmonic are 1.8 x 10$^{-4}$ and 1.5 x 10$^{-4}$.

\begin{center}
\tabcaption{ \label{tab:simu-para}  Electron and radiation parameters used in the FEL study of long modulator case.}
\footnotesize
\begin{tabular*}{82mm}{l@{\extracolsep{\fill}}cc}
\toprule
 Parameter & Value  & Unit   \\
\hline
Electron Beam &   &    \\
Energy & 4.3 & GeV \\
Energy spread & 1.0 & MeV\hphantom{0} \\
Peak current & 2.5 & kA\hphantom{0}  \\
Bunch length (full width) & 30 & fs \\
Normalized emittance & 0.5 & mm-mrad\hphantom{0} \\
Radiation & & \\
Undulator period & 3 & cm \\
$U_S$ undulator length  & 19.8 & m\\
$U_M$ undulator length (2nd/3rd) & 13/14 & m \\
$U_R$ undulator length (2nd/3rd) & 25/33 & m\\
Mono. central wavelength & 1.52 & nm \\
Mono. resolving power & 5000 & \hphantom{0} \\
Mono. power efficiency & 0.02 & \hphantom{0} \\
Bypass chicane $R_{56}$ & -0.4 & mm\\
Disp. chicane $R_{56}$ (2nd/3rd) & -0.56/-0.46 & $\mu$m\\
FEL wavelength (2nd/3rd) & 0.76/0.51 & nm\\
\bottomrule
\end{tabular*}
\vspace{0mm}
\end{center}

\begin{center}
\includegraphics[width=0.48\textwidth]{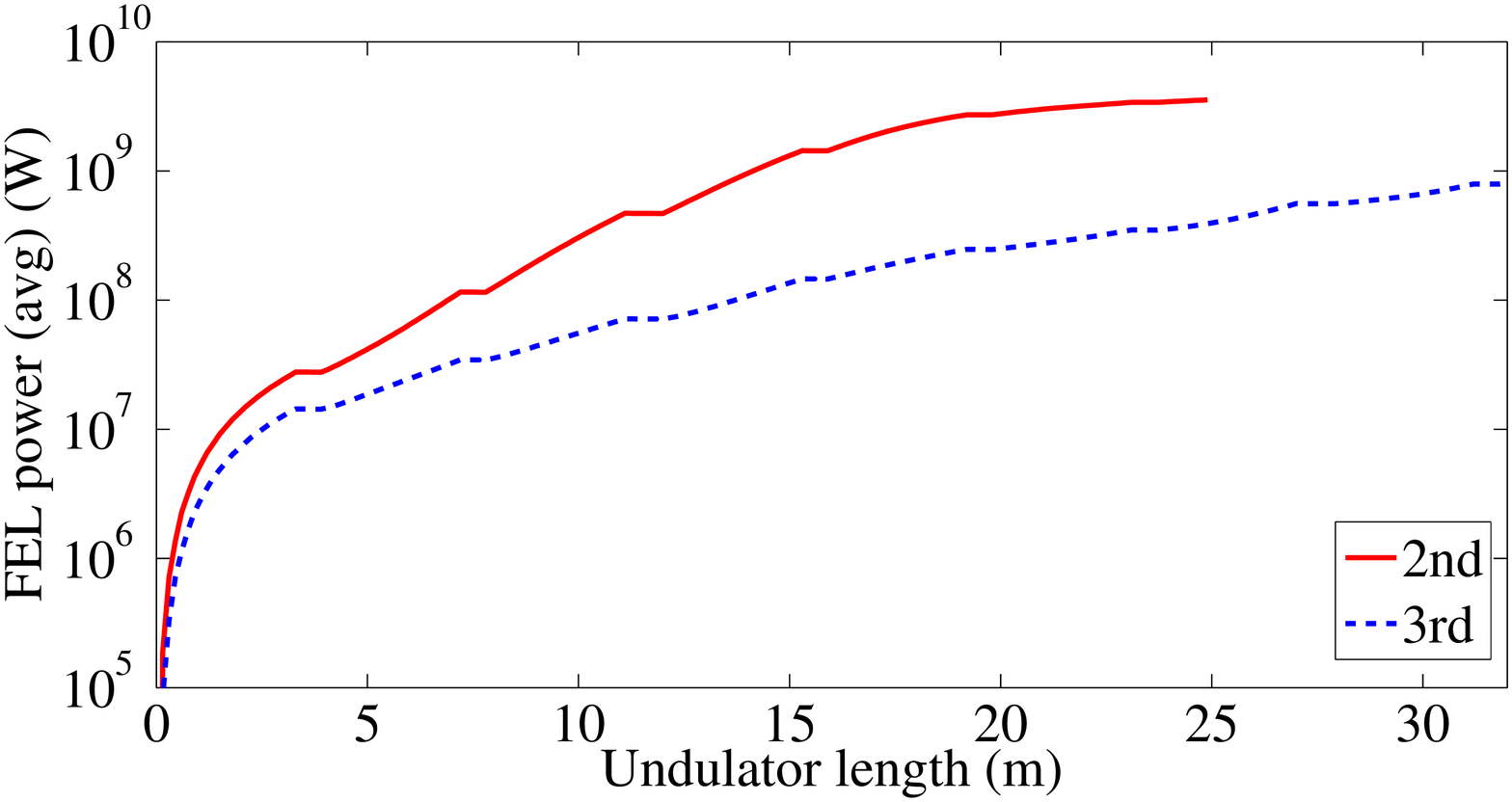}
\figcaption{\label{fig:fel:pow-vs-z}Average power evolution in the $U_R$ undulator in the long modulator case.}
\end{center}
\begin{center}
\includegraphics[width=0.5\textwidth]{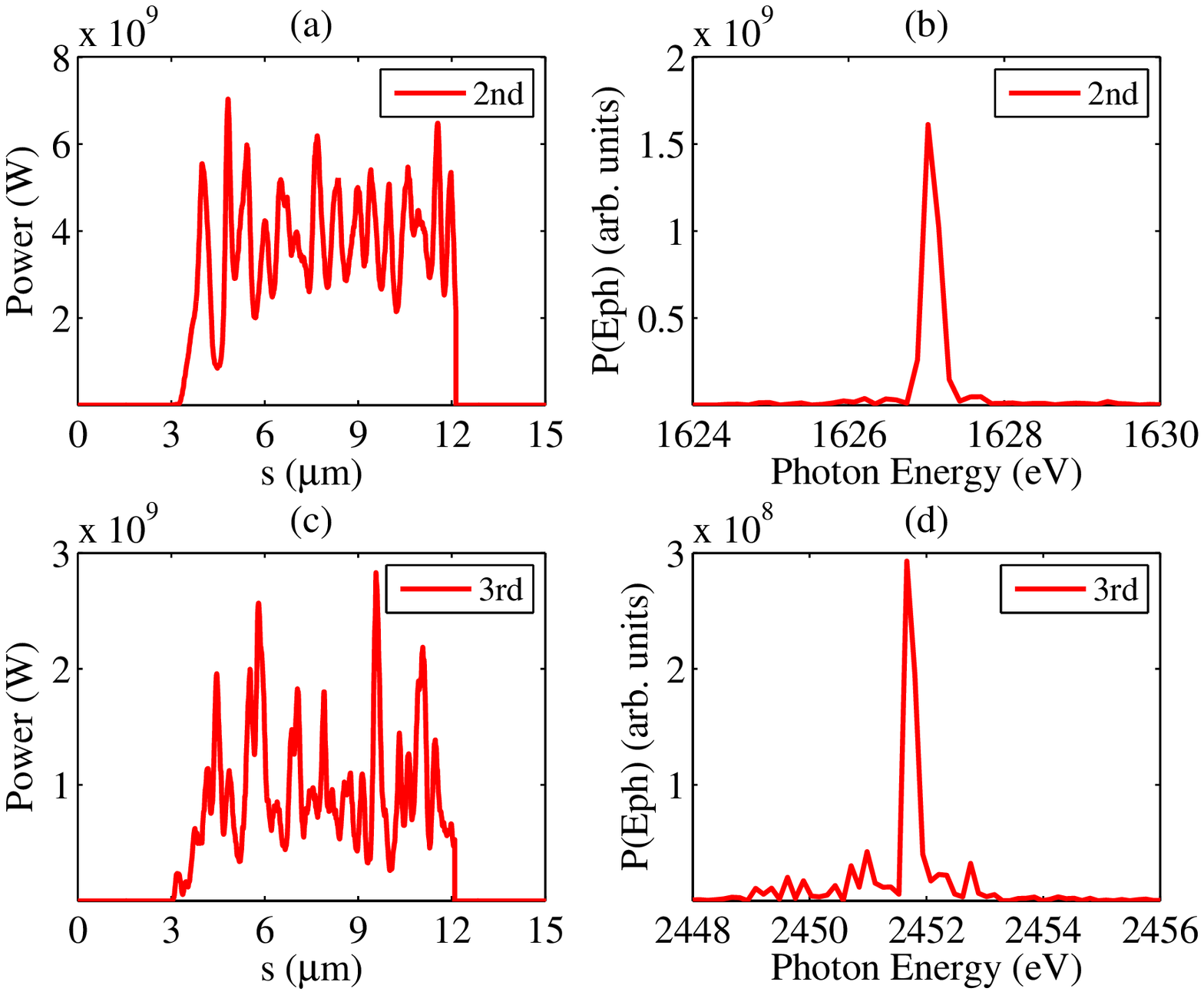}
\figcaption{\label{fig:fel:pas:1}The FEL output power profile and spectrum of the self-seeding HGHG scheme in the long modulator case.}
\end{center}

\section{FEL simulation of ``fresh bunch'' case}\label{sec:simulation:fresh-bun}

To generate the desired seed after the X-ray monochromator, a 60 fs long electron bunch is used, which is twice of that used in the long modulator case. Other electron parameters are kept unchanged, as in Table~\ref{tab:simu-para} .  Besides, the SASE undualator ($U_S$) and the X-ray monochromator remain the same as the long modulator case.

\begin{center}
\includegraphics[width=0.22\textwidth]{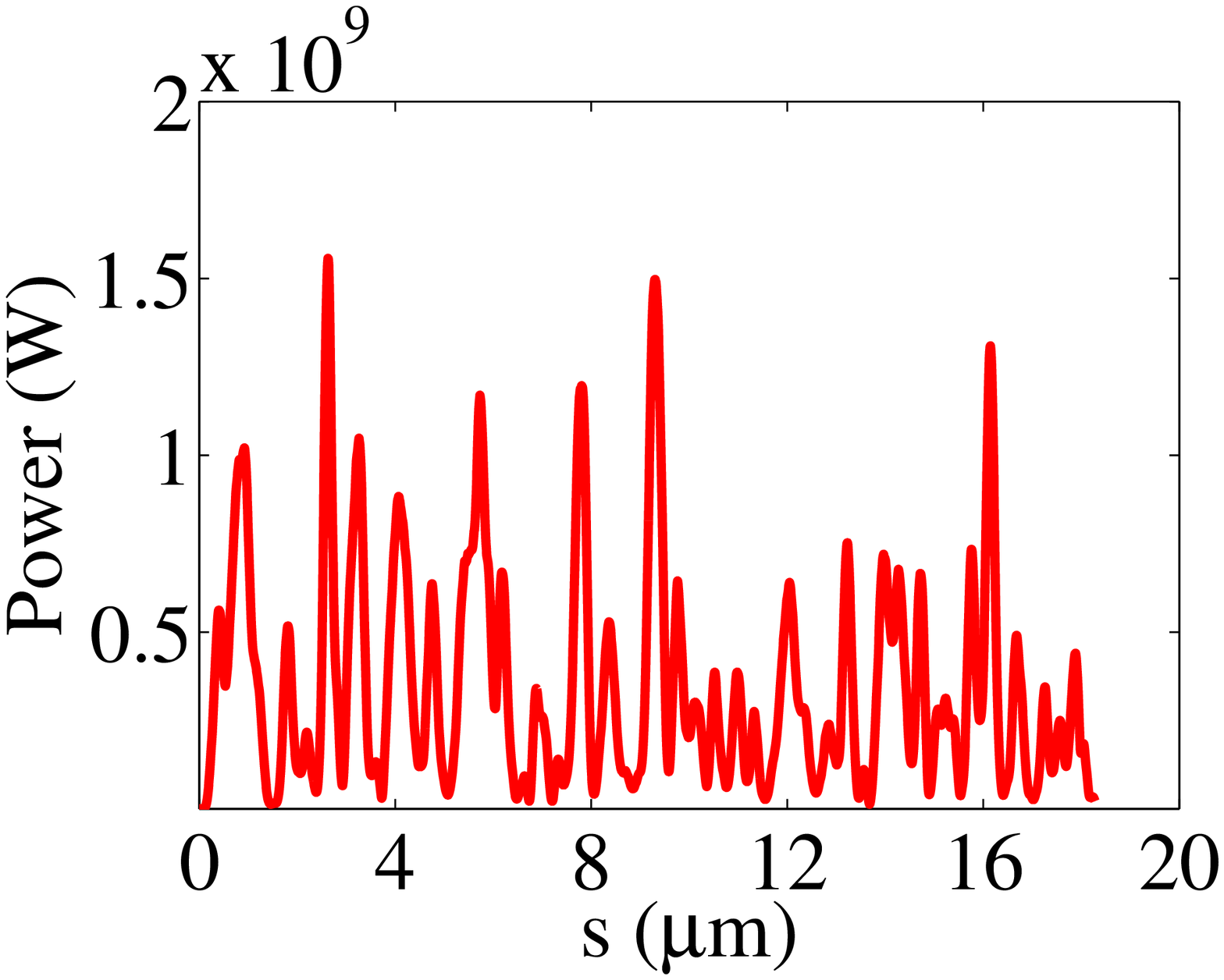}
\includegraphics[width=0.22\textwidth]{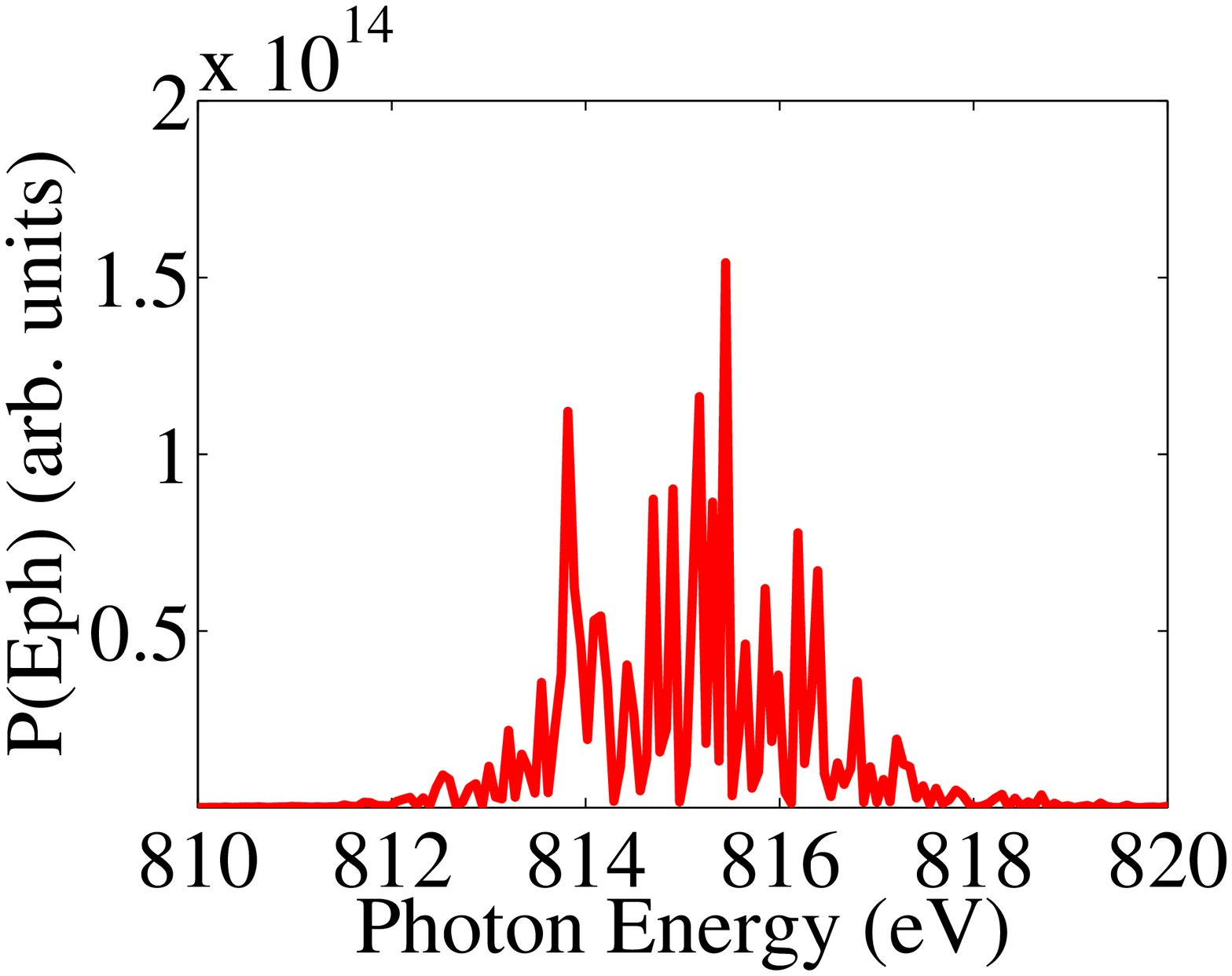}\\
\includegraphics[width=0.22\textwidth]{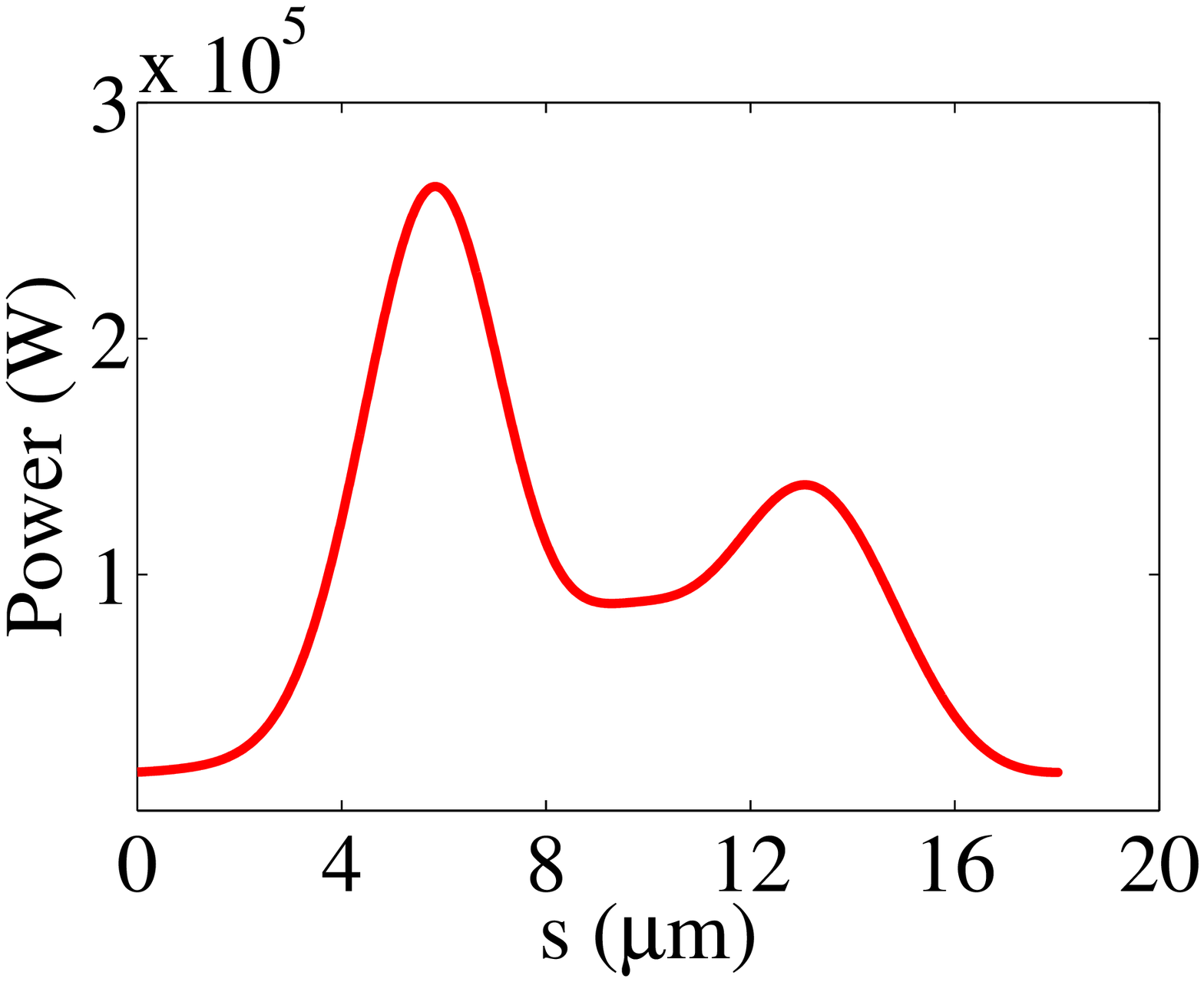}
\includegraphics[width=0.22\textwidth]{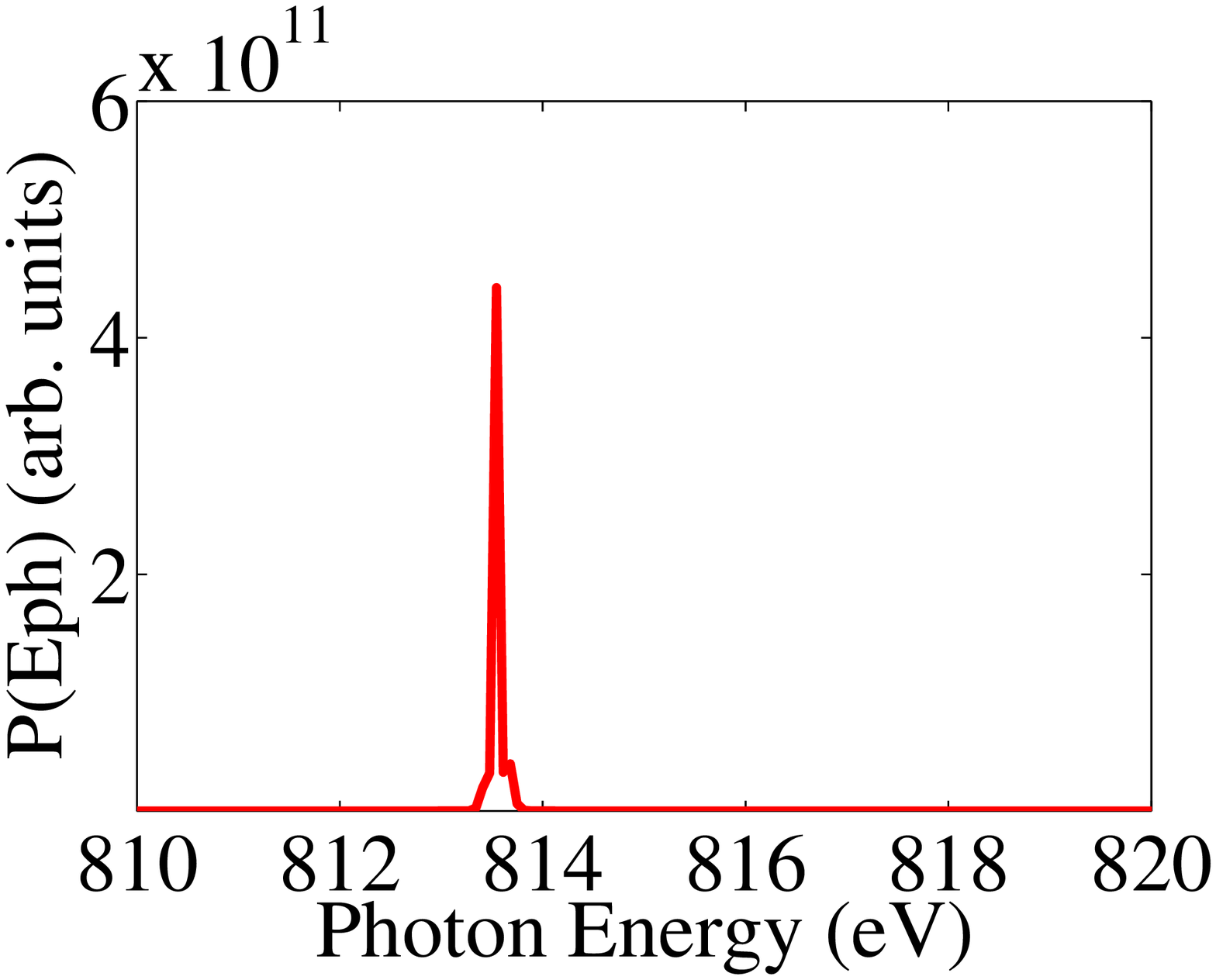}
\figcaption{\label{fig:seed:2} FEL power profiles (left) and spectra (right) at the exit of $U_S$ undulator (top) and after monochromator (bottom) in the ``fresh bunch'' case.}
\end{center}

An example of the FEL power profiles and power spectra after the SASE undulator and X-ray monochromator is shown in Fig.~\ref{fig:seed:2}. As shown in the figure, the temporal profile of the seed has double spikes. The head spike of the seed is then aligned with the tail part of the electron bunch at the entrance of the amplifying undulator ($U_A$) by fine-tuning the bypass chicane ($C_B$)  at $\mu$m level. The seed radiation copropagates with the electron bunch in the undulator and gets amplified. The length of $U_A$ undulator is chosen based on the consideration that the seed laser power can be amplified as high as possible while keeping the head part of the electron bunch fresh. In this case, the $U_A$ undulator length is 11 m. The evolution of the seed radiation pulse and electron bunch in $U_A$ is shown in Fig.~\ref{fig:fel:evol}. It can be seen from the figure that, while the peak power of the radiation is amplified to about 100 MW in the $U_A$ undulator, the head electrons in the bunch do not get disturbed. It is obvious that the energy spread of the tail electrons becomes larger, however, it is also worth noting that the energy spread increasement is not significant herein.

\begin{center}
\includegraphics[width=0.48\textwidth]{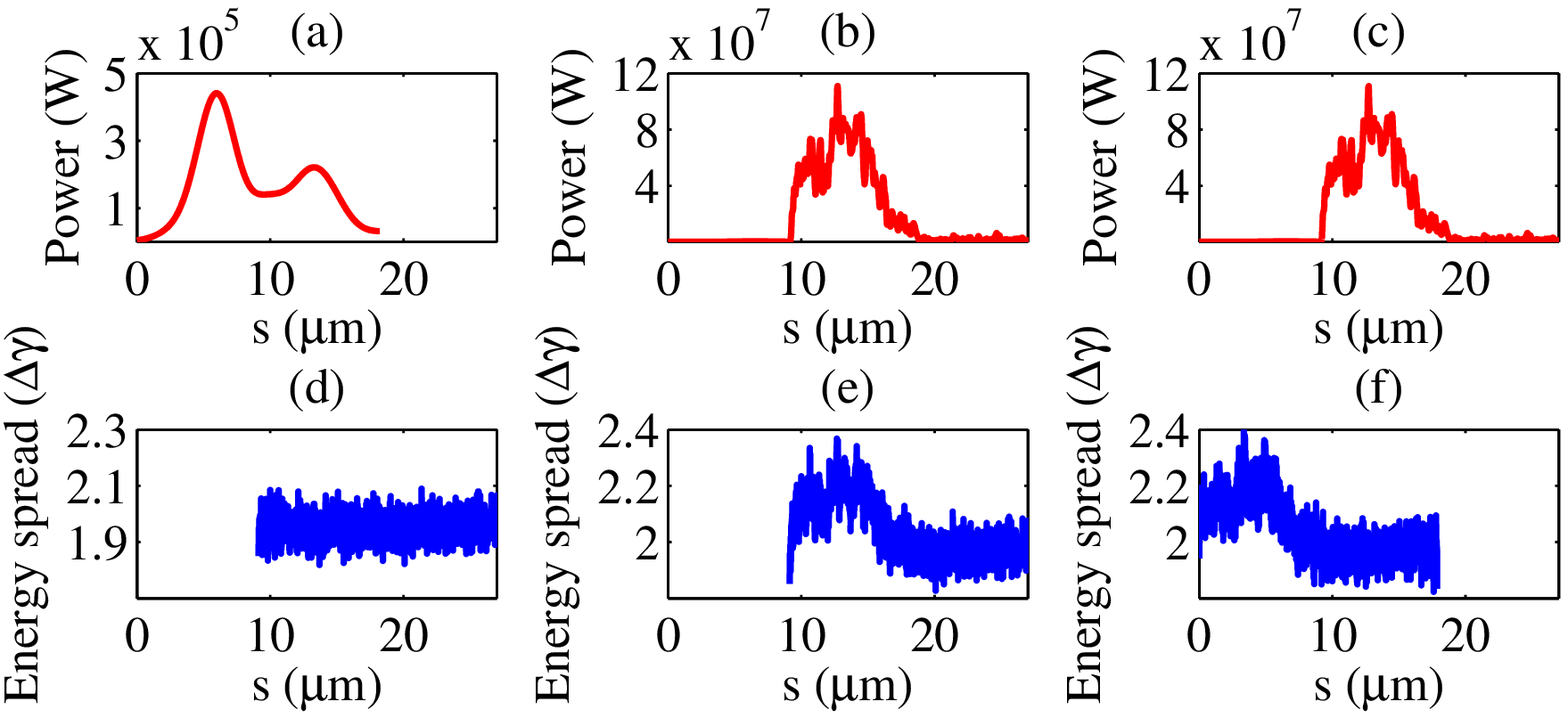}
\figcaption{\label{fig:fel:evol}  Simulated radiation power (top) and electron beam (bottom) evolution at the entrance to $U_A$ (left), at the exit of  $U_A$ (middle) and at the entrance to $U_{M2}$ (right) in the ``fresh bunch'' case.}
\end{center}

After the $U_A$ undulator, the electron bunch is delayed by the $C_{B2}$ chicane, and the head part electrons are aligned with the seed radiation in the $U_{M2}$ modulation undulator. The length of $U_{M2}$ undulator is optimized according to Eq.~(\ref{eq:opt-mod-amp}). For the 2nd and 3rd harmonic generation, the undulator length is chosen to be 2.5 m and 3.5 m, respectively.
The dispersion chicane $C_D$ in this ``fresh bunch'' case has an $R_{56}$ of 0.50 $\mu m$ and 0.42 $\mu m$ for the 2nd and 3rd harmonic generation, respectively, according to Eq.~(\ref{eq:R56}).

Figure~\ref{fig:fel:pow-vs-z-2} shows the evolution of the FEL power of the 2nd and 3rd harmonic radiation along the radiation undulator. The length of the radiation undulator $U_R$ is 20 m and 28 m, respectively. As shown in the figure, the 2nd harmonic radiation at the exit of the radiation undulator is about 7.8 GW, and the 3rd harmonic radiation with a power of about 1.9 GW. 
\begin{center}
\includegraphics[width=0.48\textwidth]{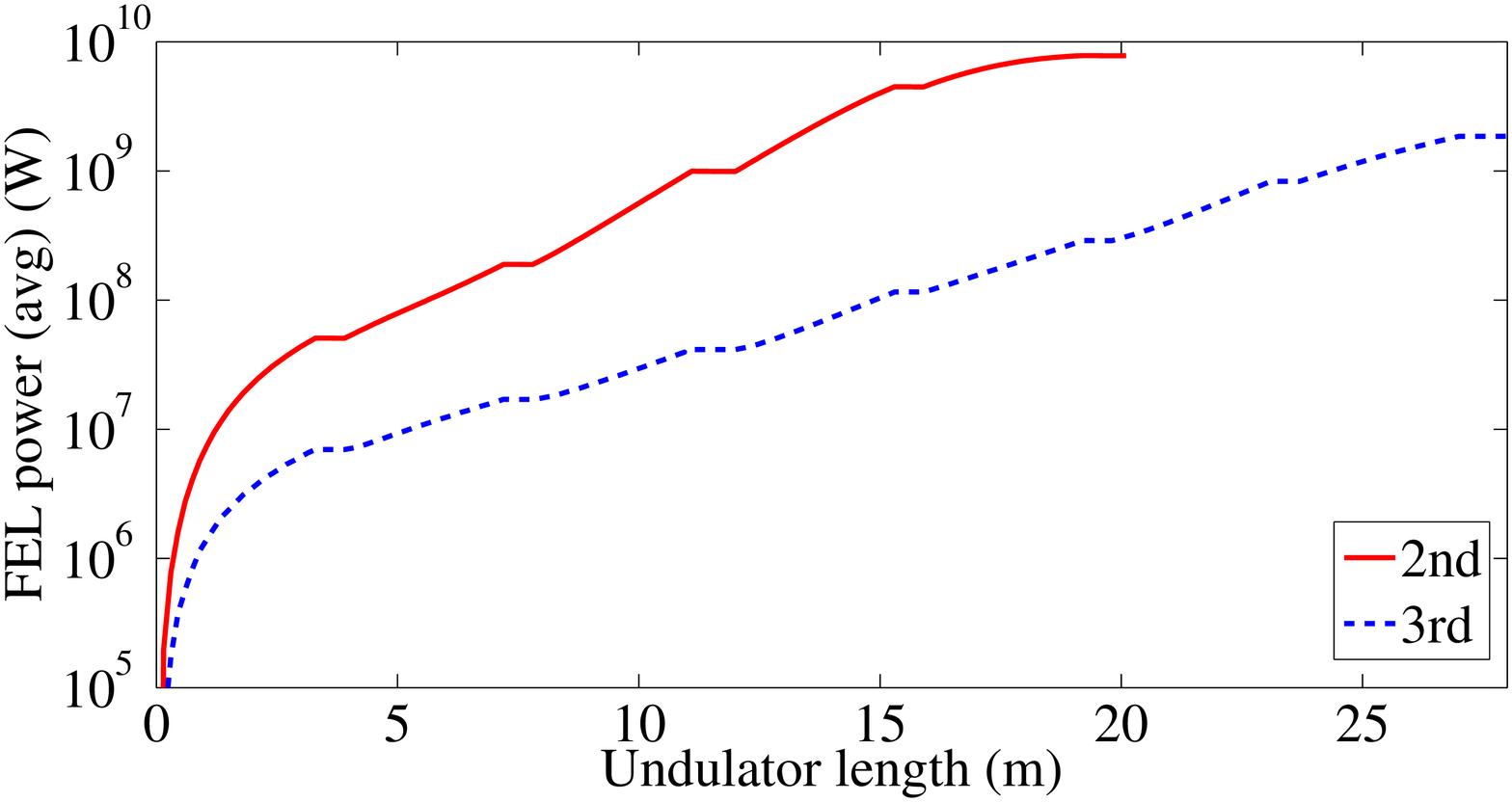}
\figcaption{\label{fig:fel:pow-vs-z-2}Average power evolution in the $U_R$ undulator in the ``fresh bunch'' case.}
\end{center}

The output power profile and spectra of the 2nd and 3rd harmonic radiation at the exit of the radiation undulator are shown in figure~\ref{fig:fel:pas:2}, which demonstrate that the normalized spectral width (FWHM) of the 2nd and 3rd harmonic are 1.6 x 10$^{-4}$ and 1.4 x 10$^{-4}$.

\begin{center}
\includegraphics[width=0.5\textwidth]{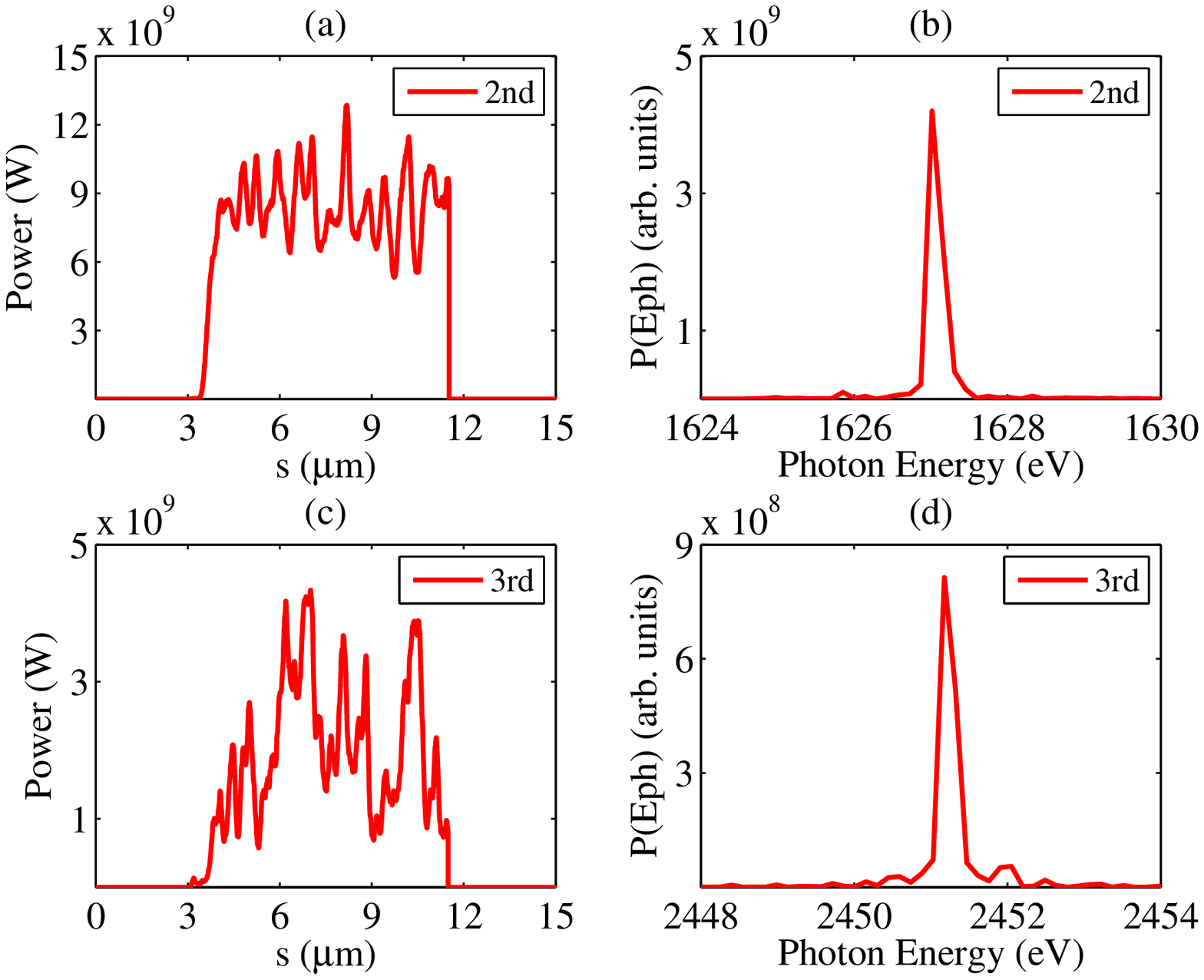}
\figcaption{\label{fig:fel:pas:2}The output power profile and spectrum of the self-seeding HGHG scheme in the ``fresh bunch'' case.}
\end{center}


To analyze the statistical fluctuation on the radiation pulse, a series of 10 separate GENESIS runs have been performed, distinguished by different
random shot noise initialization of the SASE radiation in the $U_S$ undulator. The statistical average output power for the 2nd and 3rd harmonic in the long modulator case are 2.3 GW and 0.7 GW , while that in the fresh bunch case are 5.6 GW and 1.7 GW. The shot-to-shot output power fluctuation of the 2nd and 3rd harmonic are 33.27\% (RMS) and 30.40\% (RMS) in the long modulator case£¬as well as 36.16\% (RMS) and 38.98\% (RMS) in the fresh bunch case.


\section{Conclusion}

 In this paper, we proposed a self-seeded HGHG scheme, which may be an attractive way to extend regular self-seeded FEL to shorter wavelength, especially within the photon energy range from 2 keV to 4.5 keV, which is difficult to be achieved due to a lack of monochromator materials. Besides, this method is also applicable to extend the hard X-ray self-seeding with crystals to even shorter wavelength. We use parameters based on the SXSS FEL at LCLS to simulate the scheme in two cases. In the long modulator case, we obtained 2nd harmonic with FEL power 2.3 GW and normalized spectral width (FWHM) 1.8 x 10$^{-4}$, 3rd harmonic with FEL power 0.7 GW and normalized spectral width 1.5 x 10$^{-4}$. While in the ``fresh bunch'' case, we acquired 2nd harmonic with FEL power 5.6 GW and normalized spectral width 1.6 x 10$^{-4}$,  3rd harmonic with FEL power 1.7 GW and normalized spectral width 1.4 x 10$^{-4}$.

 Note in these demonstration examples, we chose a fundamental energy at 810 eV to calculate the second and third harmonic performance. The soft x-ray selfseeding setup can work at 1.5 keV fundamental energy with the same monochromator, so we can scale up to reach 4.5 keV at the third harmonic to cover the energy gap we discussed earlier. Also we used a conserved energy spread 1 MeV rms based on the present LCLS operation parameters. A lower energy spread is possible based on LCLS-II design studies which should help the harmonic performance. This will be investigated in the future.

\end{multicols}

\vspace{-1mm}
\centerline{\rule{80mm}{0.1pt}}
\vspace{2mm}

\begin{multicols}{2}

\end{multicols}

\clearpage
\end{CJK*}
\end{document}